\documentclass[aps,prd,twocolumn,preprintnumbers,
floatfix,nofootinbib]{revtex4}

\usepackage{graphicx,amssymb,amstext,mathrsfs,dsfont,amsfonts}
\usepackage{eucal,wrapfig,boxedminipage,setspace,subfigure,epsfig}
\usepackage{color,latexsym,url}
\usepackage[colorlinks,hyperindex]{hyperref}
    \hypersetup
    {
        colorlinks,%
        citecolor=red,%
        linkcolor=blue,%
        urlcolor=black,%
    }

\def\a  {\alpha}       \def\b  {\beta}         \def\g  {\gamma}
\def\G  {\Gamma}       \def\d  {\delta}        
     \def\ve {\varepsilon}   
\def\l  {\lambda}

   \def\w  {\omega}

 \newcommand{\call}{\mbox{${\cal L}$}}

 \newcommand{\calr}{\mbox{${\cal R}$}}

\def\IR{{\hbox{{\rm I}\kern-.2em\hbox{\rm R}}}}
\def\IB{{\hbox{{\rm I}\kern-.2em\hbox{\rm B}}}}
\def\IN{{\hbox{{\rm I}\kern-.2em\hbox{\rm N}}}}
\def\IC{\,\,{\hbox{{\rm I}\kern-.59em\hbox{\bf C}}}}
\def\IZ{{\hbox{{\rm Z}\kern-.4em\hbox{\rm Z}}}}
\def\IP{{\hbox{{\rm I}\kern-.2em\hbox{\rm P}}}}
\def\IH{{\hbox{{\rm I}\kern-.4em\hbox{\rm H}}}}
\def\ID{{\hbox{{\rm I}\kern-.2em\hbox{\rm D}}}}

\def\be{\begin{equation}}
\def\ee{\end{equation}}
\def\ba{\begin{eqnarray}}
\def\ea{\end{eqnarray}}

\newcommand{\inv}[1]{\frac{1}{#1}}

\def\ra{\rightarrow}

\def\del{\partial}

\def\det{{\rm det}}

\def\nn{\nonumber}
\def\ea{{\it et al}. }
\newcommand{\eg}{{\it e.g.$\, $}}
\newcommand{\ie}{{\it i.e.$\, $}}

\newcommand{\wt}{\widetilde}

\newcommand{\beq}{\begin{equation}}
\newcommand{\eeq}{\end{equation}}
\newcommand{\bea}{\begin{eqnarray}}
\newcommand{\eea}{\end{eqnarray}}

\newcommand{\guv}{{g_{\mathrm{UV}}}}
\newcommand{\guvt}{{g^2_{\mathrm{UV}}}}

\begin{document}

\newcommand\sect[1]{\emph{#1}---}

\preprint{
\begin{minipage}[t]{3in}
\begin{flushright} SHEP-10-31
\\[30pt]
\hphantom{.}
\end{flushright}
\end{minipage}
}

\title{Holography of a Composite Inflaton}

\author{Nick Evans}
\email{evans@soton.ac.uk}
\author{James French}
\email{james.french@soton.ac.uk}
\author{Keun-Young Kim}
\email{k.kim@soton.ac.uk}

\affiliation{ School of Physics and Astronomy, University of
Southampton, Southampton, SO17 1BJ, UK}

\begin{abstract}
We study the time evolution of a brane construction that is
holographically dual to a strongly coupled gauge theory that
dynamically breaks a global symmetry through the generation of an
effective composite Higgs vev. The D3/D7 system with a background
magnetic field or non-trivial gauge coupling (dilaton) profile
displays the symmetry breaking. We study motion of the D7 brane in
the background of the D3 branes. For small field inflation in the field theory the
effective Higgs vev rolls from zero to the true vacuum value. We
study what phenomenological dilaton profile generates the slow rolling needed,
hence learning how the strongly coupled gauge theory's
coupling must run. We note that evolution of our configuration in the holographic
direction, representing the phyiscs of the strong interactions,
 can provide additional slowing of the roll time. Inflation seems to be
favoured if the coupling changes by only a small amount or very gently.
We speculate on how such a scenario could be
realized away from N=4 gauge theory, for example, in a walking gauge
theory.

\noindent

\end{abstract}

\maketitle

\section{Introduction}\vspace{-0.5cm}

In recent years the AdS/CFT Correspondence
\cite{Malda,Witten:1998qj,Witten:1998b,Gubser:1998bc} has emerged
as a very powerful tool for studying strong coupling problems. It
provides a weakly coupled gravitational/string description of
strongly coupled gauge dynamics allowing computation of many
aspects of the dynamics. Originally the duality was proposed for
the ${\cal N}=4$ super Yang Mills theory but it has been widely
expanded to other theories over the past ten years. Amongst those
theories for which gravity duals exist are several examples of
strongly coupled gauge theories with fundamental quark fields and
which break chiral symmetries by the formation of a quark
condensate as QCD does
\cite{Babington,Kruczenski:2003uq,Ghoroku:2004sp,Alvares:2009hv}.
In this paper we wish to use these tools to make a first study of
time dependence in such symmetry breaking theories. We expect
strongly coupled gauge theories to undergo phase transitions
between a high temperature chirally symmetric phase and a low
temperature chiral symmetry breaking phase. Holography should for
the first time allow us to analyse the evolution of such theories
between these different phases. The simplest example of where such
a transition might play an important role in the evolution of our
Universe is in inflation and in this paper we will concentrate on
that possibility.

Inflation (see for example \cite{Lyth:1998xn}) is now a key part
of the standard cosmological model of the Universe, supported by
many pieces of astrophysical data. The usual description involves
one or more scalar fields slow-rolling from an unstable point in a
potential to the true vacuum. The origin for such a scalar field
remains unclear though. Indeed, until evidence for supersymmetry
is found in nature, fundamental scalar fields are formally
unnatural in field theory as a result of the hierarchy problem (radiative corrections cause their mass
to naturally grow to the Planck scale).
This need be no obstacle to the paradigm though since many
strongly coupled systems have scalar order parameters describing
their dynamics that could play the same role. For example, in QCD
the gauge interactions dynamically generate a quark bilinear
condensate ($\langle \bar{q} q \rangle$) that breaks chiral
symmetries - one may think of that bilinear as the expectation
value of a composite scalar field. The main reason to work with
fundamental scalars in cosmology is simply that we have had no
tools to study strong coupling problems such as the condensation
in QCD. The AdS/CFT Correspondence for the first time allows us to
analyse just such a scenario though.

We will concentrate on a particular duality which we believe to be
the simplest example of holography with fundamental quark fields
\cite{Polchinski,Bertolini:2001qa,Karch,Mateos,Erdmenger:2007cm}.
We wish to stress that we do not consider the specific degrees of
freedom of the theory too crucial - it is some strongly coupled
gauge theory that generates quark condensates. We hope, in the
spirit of AdS/QCD models\cite{Erlich:2005qh,Da Rold:2005zs}, that
it reflects broad aspects of many strongly coupled systems. The
specific gauge theory is constructed from the D3/D7 system\footnote{The
D3/D7 system has been used to construct an inflation model in \cite{d3d7inflation} and
subsequent literature. The motivation in those models is rather different
since they are not holographic descriptions of a strongly coupled gauge theory
but instead assumed to describe weakly coupled fields
with the extra dimensions compactified. The dilaton profiles and brane motions
we will consider are very different being inspired by the gauge
duality and the lessons we will seek to extract are for strongly coupled
gauge dynamics. Another interesting AdS/CFT approach for a strongly coupled inflaton was reported in \cite{Xingang} }
in type
IIB string theory which we will describe in detail below. The
theory is the large $N_c$ ${\cal N}=4$ $U(N_c)$ gauge theory with
a small number of quark hypermultiplets. We will work in the
quenched approximation \cite{Karch}(appropriate when $N_f \ll
N_c$) which on the gravity dual side corresponds to treating the
D7 branes as probes in the AdS metric generated by the D3 branes.
There is a $U(1)$ symmetry (a remnant of the $SU(4)$ R-symmetry of
the ${\cal N}=4$ theory) which is broken when a quark condensate
forms \cite{Babington}. Several mechanisms for triggering this
condensation have been explored. The cleanest is when a background
magnetic field is introduced \cite{Magnetic,Evans:2010iy}. Running
of the coupling also causes quark condensation as has been shown
in back-reacted dilaton flow geometries
\cite{Babington,Ghoroku:2004sp}and models with a
phenomenologically imposed dilaton profile\cite{Alvares:2009hv}.
The quark condensate can be determined in these models and an
effective IR quark mass is generated. The theories display a
massless pion-like Goldstone field and a massive sigma field
(since we are at large $N_c$ it is stable) that is the effective
Higgs particle.

We will seek to learn about a strongly coupled counter part to a
``new inflation" or ``small field inflation" model \cite{newinflation}. One assumes a high
temperature phase where the effective Higgs scalar's vev is zero
and a low temperature phase where the vev is non-zero. A second
order phase transition should connect these phases so that the
vacuum is left at the unstable zero vev point of the low
temperature theory and then rolls to the true vacuum. For
inflation the potential needs to be very flat near the origin so
the roll takes a long time. Of course, other possible
inflationary scenarios exist with multiple scalar fields or where
the scalar expectation value is initially large relative to the
vacuum value and it would be interesting to investigate these
ideas holographically too in the future.

In principle one could imagine taking a holographic model and
tracking it through the finite temperature transition. On the
gravity side temperature is introduced through a black hole in the
AdS geometry. The D3/D7 model we investigate has been shown to
display both first order (eg with a magnetic field
\cite{Magnetic,Evans:2010iy}) and second order (eg with a magnetic
field and chemical potential \cite{Evans:2010iy}) symmetry
restoration transitions. Ideally one would work in a time
dependent background describing a shrinking black hole. In fact
such geometries are known \cite{Janik:2010we} and a first study of
D7 branes moving slowly near a potential minimum in those
geometries can be found in \cite{Grosse:2007ty}. Dealing with such
branes when they touch the black hole is hard though - we hope to
study this problem in a future publication \cite{Evans:2010xs}.

Our goal in this paper is more limited though and similar to the
usual simplest field theory analysis of inflation models. We will
assume that below the phase transition the theory is well
described by the $T=0$ theory. We will invoke initial
conditions that place the vacuum in the symmetric phase and then
watch it roll to the true vacuum that breaks the symmetry. 
These initial conditions assume the existence of a second order
transition.\footnote{Also see \cite{Das} for a time-dependent 
D5 brane embedding dynamics in the context of a quantum quench.
Non-equilibrium dynamics has been studied also 
in holographic superconductor~\cite{Murata}
} Our first results are numerical simulations of this
roll in the theory where a magnetic field is inducing the symmetry
breaking. As we said above this theory actually has a first order
thermal transition rather than a second order one - the scenario
serves to demonstrate the formalism though. The results would be
relevant to the phase transition within a highly supercooled
bubble.  Here there is no fine tuning to make the potential
particularly flat so the model would make a poor
inflation model. It serves as a proof of the numerical techniques
we use to study the problem and it is impressive that a relatively
simple computation can track such a transition in a strongly
coupled gauge theory.

One does not imagine that a magnetic field induced symmetry
breaking is likely in a realistic model of inflation. We therefore
turn to a more phenomenological model in which we embed the D7
brane in an AdS geometry but with a hand chosen and non-back
reacted dilaton profile. Whilst this is not a completely kosher
string dual we hope that it allows us to understand broadly the
dependence of a strongly coupled theory's dependence on its
running coupling. We used a similar approximation in
\cite{Alvares:2009hv} to study walking gauge dynamics and found
agreement between the gravity dual description and the usual
expectations for the coupling dependence of the chiral symmetry
breaking. The model is sufficiently simple that we can understand
how the running of the gauge coupling affects the potential for
the effective Higgs mode. We attempt to engineer models with a
slow roll between the symmetric and broken phases. We also show that part of
our success in slowing the roll time is due to additional
dynamics in the holographic radial coordinate of the dual description.
This dynamics reflects the strong coupling dynamics of the gauge theory
which seems under some circumstacnes to favour inflation. From this
analysis we conclude that gauge theories displaying a small increase
in the gauge coupling, preferably over a wide energy regime, 
are liable to make good inflation models.

Finally we briefly speculate on how IR conformal fixed points may well
be common in strongly coupled gauge dynamics and suggest a wider
set of theories that might display running of the coupling like that
we find gives inflation. Obvious examples are walking gauge theories.

\section{Inflation}

Let us very briefly review the standard inflation model
\cite{Lyth:1998xn}.
Based on the cosmological principle
we assume the  Friedmann-Robertson-Walker metric of the form
\beq ds^2 = -dt^2 + a(t)^2 \left(
{dr^2 \over 1- kr^2} +
r^2 ( d\theta^2 + \sin^2\theta d\phi^2) \right) \ , \eeq
and the matter energy momentum tensor of a perfect fluid
$T^\mu_\nu=$ diag($-\ve,p,p,p)$. For simplicity we will
work in the $k=0$ case.
Then the general Einstein equations describing the expansion of the Universe are
reduced to two equations:
\begin{eqnarray}
  && H^2 = \frac{1}{3}\ve \  \qquad \ \mathrm{:Friedmann\ equation} \ ,  \label{Friedmann} \\
  && \dot{\ve} +  3H(\ve+p) = 0 \  \ \ \mathrm{:Fluid\ equation} \ , \label{Fluid}\
\end{eqnarray}
where $H(t) \equiv \dot{a} / a$ is the Hubble parameter and the
scale factor $a(t)$ is readily determined from $H(t)$ as
\beq a(t) = a(0) \exp\left(\int^t_0 H(t') dt'\right) \ . \label{sH} \eeq

Equations (\ref{Friedmann}) and (\ref{Fluid}) can be combined to form 
a so called ``acceleration equation''
\begin{eqnarray}
  \frac{\ddot{a}}{a} = -\frac{1}{6}(\ve+3p) \ , \label{accell}
\end{eqnarray}
from which the condition for inflation can be expressed in terms of
the energy momentum tensor.
\begin{eqnarray}
  \mathrm{Inflation} \quad \Leftrightarrow \quad 
  \ddot{a} >0 \quad \Leftrightarrow \quad p < -\frac{\ve}{3} \ . \label{negativep}
\end{eqnarray}
Thus a natural question for inflation is asking what kind of matter and 
dynamics can generate sufficient negative pressure (\ref{negativep}).

For example let us consider a scalar field.

\begin{eqnarray}
  S = \int d^4x \sqrt{-g}  {1 \over 2} \calr + S_M \ ,
\end{eqnarray}
where $\calr$ is Ricci scalar and
\beq S_M = \int d^4x \sqrt{-g} \left[ - {1 \over 2}
g^{\mu \nu} \partial_\mu \varphi \partial_\nu \varphi - V(\varphi) \right] \ , \label{SM}
\eeq
in unit of $8\pi G = 8 \pi m_P^{-2} = 1$.
$m_P$ is the Plank mass.
From a scalar matter action the energy-momentum tensor
of a homogeneous field $\varphi(t)$ can be obtained as
\beq \label{rhp} \ve = - T^0_0 = {1 \over 2} \dot{\varphi}^2 + V(\varphi),
\hspace{0.5cm} p = T^i_i= {1 \over 2} \dot{\varphi}^2 - V(\varphi) \ . \label{ep1}
\eeq
The inflation condition (\ref{negativep}) is rephrased as 
\begin{eqnarray}
  p < -\frac{\ve}{3} \quad \Leftrightarrow \quad \dot{\varphi}^2 < V(\varphi) \ .
\end{eqnarray}
Whenever the kinetic term is small compared to the potential energy there will be inflation.
Then by analogy to mechanics we may say $\varphi$ is {\it rolling slowly}.

Let us apply this slow roll condition to the equations of motion 
(\ref{Friedmann}) and (\ref{Fluid}). 
\begin{eqnarray}
  && H^2 = {1 \over 3} \ve =  {1 \over 3} \left({1 \over 2} \dot{\varphi}^2 + V(\varphi) \right)\ ,
  \label{H2}  \\
  && \ddot{\varphi} + 3 H
\dot{\varphi}+{d V \over d \varphi}=0 \label{scalar} \ , \label{ddphi}
\end{eqnarray}
where the fluid equation (\ref{ddphi}) is nothing but the equation of motion for $\varphi$, which
can be obtained from (\ref{SM}).

If the scalar field stays in some part of the potential
with $\dot{\varphi}^2 \ll V$
then $H \sim V(\varphi(t))/3$ from (\ref{H2}).
Furthermore if $\ddot{\varphi}$ is sufficiently smaller than
the other terms in (\ref{ddphi}) then $V(\varphi)$ can be considered as
constant, say $V_0$, for a long time.
Thus the scale factor (\ref{sH}) in the slow rolling range reads
\beq a(t) \sim a(0) e^{Ht} =  a(0) e^{ \sqrt{\frac{8\pi V_0}{3}}\inv{m_P} t}  \ , \eeq
where we have reinserted the
Planck mass.
This is the de Sitter limit
zeroth order approximation of an inflating Universe,
which we will adopt in this paper. To have a graceful exit from inflation
the Hubble parameter must be time dependent and fall to zero at late times.
In an inflating regime the space is equivalent to de Sitter space with a cosmological
constant $\Lambda = 8\pi m_P^{-2}V_0$, where
$V_0$ is interpreted as
the vacuum energy density.
The amount of inflation is specified by the number of e-folds given by
\beq N_e \equiv \log\frac{a(t_e)}{a(0)}= H t_e =
{\sqrt{8 \pi V_0 \over 3} \frac{t_e}{m_P}} \ , \label{efold} \eeq
where $t_e$ is the time when inflation ends
(when slow roll conditions are violated)
starting from $t=0$.
One needs $N_e$ to be more than 60 phenomenologically.

Note that a larger $H$ yields a more inflationary evolution.
$H$ is nothing but a friction coefficient when we
interpret (\ref{ddphi}) as an equation for the
classical particle trajectory ($\varphi(t)$) under the potential $V$.
Thus it is natural that a larger friction induces a slower
rolling of the particle.

In this paper we will replace the scalar field sector with a
strongly coupled gauge theory. The gauge dynamics will generate a
wine bottle shaped potential with a non-zero condensate of a quark
bilinear at the minimum. We will phenomenologically adjust the
running of the gauge theory's coupling constant to control the
shape of the potential. We hope to learn how a gauge theory's
coupling must run to generate inflation. Clearly our task is to
compute the stress energy tensor ($\ve$ and $p$) of a strongly
coupled gauge theory so we can substitute them into
the Friedmann and fluid equations above.
We must find the equivalent of the Euler
Lagrange equation for the scalar field $\varphi$ in the inflating
background for the time evolution of the strongly coupled gauge
dynamics. To do this we will turn to Gauge/Gravity
duality.

\section{Holographic Description of a Strongly Coupled Gauge
Theory}

\subsection{Brane Construction}

First let us review the gravity dual description of the symmetry
breaking behaviour of our strongly coupled gauge
theory\cite{Babington,Kruczenski:2003uq,Ghoroku:2004sp,Alvares:2009hv}.
To begin with we will work in a time independent flat space (\ie
there is no inflation here and $a(t)=1$).

\begin{figure}[]
\centering
   {\includegraphics[width=7cm]{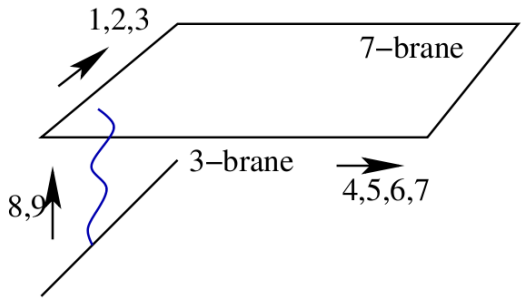}}
\caption{
           {\small A sketch of the D3/D7 construction that our model
is based on. 3-3 strings generate the gauge theory whilst 3-7 strings
are quark fields.}
           }\label{D3D7}
\end{figure}

D$p$-branes are $p$ dimensional membrane-like objects on which the
ends of open strings can be fixed. The weak coupling picture for our D3/D7
set up is shown in Fig \ref{D3D7} - there are $N$ D3 branes and the
lightest string states with both ends on that stack generate the
adjoint representation fields of the ${\cal N}=4$ gauge theory.
Strings stretched between the D3 and the D7 are the quark fields
lying in the fundamental representation of the $SU(N)$ group (they
have just one end on the D3).

In the strong coupling limit the D3 branes in this picture are
replaced by the geometry that they induce
\cite{Malda,Witten:1998qj,Witten:1998b,Gubser:1998bc}.
We will
consider a gauge theory with a holographic dual described by the
Einstein frame geometry $AdS_5 \times S^5$ 
\beq ds^2 = {1 \over
\guv} \left[{r^2 \over R^2} dx_{4}^2 + {R^2 \over r^2} \left(
d\rho^2 + \rho^2 d\Omega_3^2 + dw_5^2 + dw_6^2 \right)\right] \ , 
\eeq
where we have split the coordinates into the $x_4(x_{3+1})$ of the
gauge theory, the $\rho$ and $\Omega_3$ which will be on the D7
brane world-volume and two directions transverse to the D7,
$w_5,w_6$. The radial coordinate, $r^2 = \rho^2 + w_5^2 + w_6^2$,
corresponds to the energy scale of the gauge theory. The radius of
curvature is given by $R^4 = 4 \pi \guvt N_c \alpha'^{2}$ with
$N_c$ the number of colours.  In addition we will allow an
arbitrary dilaton as $r \rightarrow 0$ to represent the running of
the gauge theory coupling
\beq
e^{\Phi} = g^2_{\mathrm{YM}}(r^2) = \guvt ~\beta(r^2) \ , \label{beta}
\eeq
where the function $\beta(r) \rightarrow 1$ as
$r \rightarrow \infty$. The $r \rightarrow \infty$ limit of this
theory is dual to the ${\cal N}=4$ super Yang-Mills theory and
$\guvt$ is the constant large $r$ asymptotic value of the gauge
coupling.

We will introduce a single D7 brane probe~\cite{Karch} into the
geometry to include quarks - by treating the D7 as a probe we are
working in a quenched approximation although we can reintroduce
some aspects of quark loops through the running coupling's form if
we wish (or know how). This system has a U(1) symmetry acting on
the quarks, corresponding to rotations in the $w_5$-$w_6$ plane,
which will be broken by the formation of a quark condensate.

In the true vacuum at zero temperature ($T=0$) the brane will be static. We must find
the D7 embedding function \eg $w_5(\rho) \equiv L(\rho)$ with $w_6=0$.
The DBI action in Einstein frame is given by
\beq
\begin{array}{ccl}
S_{D7} & = & -T_7 \int d^8\xi e^\Phi  \sqrt{- \det P[G]_{ab}}\\ &&\\
&=&  -\wt{T}_7 \int d^4x~ d \rho ~ \rho^3 \beta \sqrt{1 +
L'^2} \ , \end{array} \label{Sd7}
\eeq
where $P[G]_{ab}$ is the pull back of the background metric onto
the D7 and $L' \equiv \del_\rho L(\rho)$. $T_7 = 1/(2 \pi)^7
\alpha'^{4}$ and $\wt{T}_7 = 2 \pi^2 T_7/ \guvt$ when we have
integrated over the 3-sphere on the D7. The equation of motion for
the embedding function is therefore
\beq \label{embed}
\partial_\rho \left[ {\beta \rho^3
L' \over \sqrt{1+ L'^2}}\right] - 2 L \, \rho^3
\sqrt{1+ L'^2} {\partial \beta(r^2) \over \partial
r^2} = 0 \ . \eeq
The UV asymptotic of this equation, provided the
dilaton returns to a constant so the UV dual is the ${\cal N}=4$
super Yang-Mills theory, has solutions of the form
\beq
\label{asy} L = m + {c \over \rho^2} +... \ , \eeq
where we can
interpret $m$ as the quark mass ($m_q = m/2 \pi \alpha'$) and $c$
is proportional to the quark condensate.

The embedding equation (\ref{embed}) clearly has regular solutions
$L=m$ when $g^2_{YM}$ is independent of $r$ - the flat embeddings
of the ${\cal N}=2$ theory \cite{Karch,Mateos}. Equally clearly if
$\partial \beta(r^2) / \partial r^2$ is none trivial in $L$ then
the second term in (\ref{embed}) will not vanish for a flat
embedding.

There is always a solution $L=0$ which corresponds to a massless
quark with zero quark condensate ($c=0$). In the high $T$ phase this
is the true vacuum. In the symmetry breaking low $T$ geometry this
configuration is a local maximum in the potential.

Note that in the particular case when
\beq \beta = \sqrt{1 + {B^2
\over (\rho^2 + L^2)^2}} \ , 
\eeq
the DBI action for the D7 brane is that of the D3/D7 system with a
background magnetic field $B$. This model has been extensively
studied in \cite{Magnetic,Evans:2010iy}. The action manifestly
grows as one approaches $L = \rho = 0$ so the D7 brane is repelled
from that point. In Fig 2 we plot the D7 embedding in the magnetic
field case.

An interesting phenomenological case is to consider a gauge
coupling running with a step of the form \beq
\label{coupling}\beta =  A +1 - A \tanh\left[ \Gamma (r - \lambda)
\right] \ . \eeq  Of course in this case the geometry is not back
reacted to the dilaton and the model is a phenomenological one in
the spirit of AdS/QCD. This form introduces conformal symmetry
breaking at the scale $\Lambda = \lambda/2 \pi \alpha'$ which
triggers chiral symmetry breaking. The parameter $A$ determines
the increase in the coupling across the step. If the coupling is
larger near the origin then again the D7 brane will be repelled
from the origin.  The parameter $\Gamma$ spreads the increase in
the coupling over a region in $r$ of order $\Gamma^{-1}$ in size -
the effect of widening the step is to enhance the large $\rho$
tail of the D7 embedding.

\begin{figure}[]
\centering
    {\includegraphics[width=7cm]{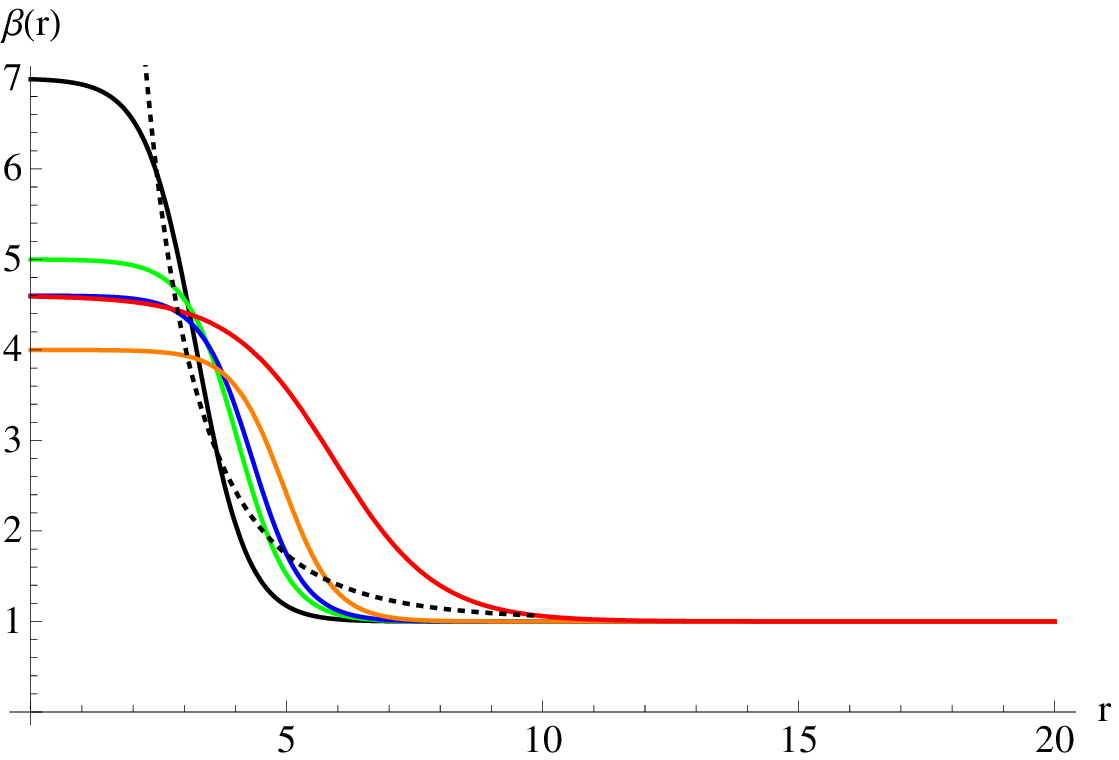}}
    {\includegraphics[width=7cm]{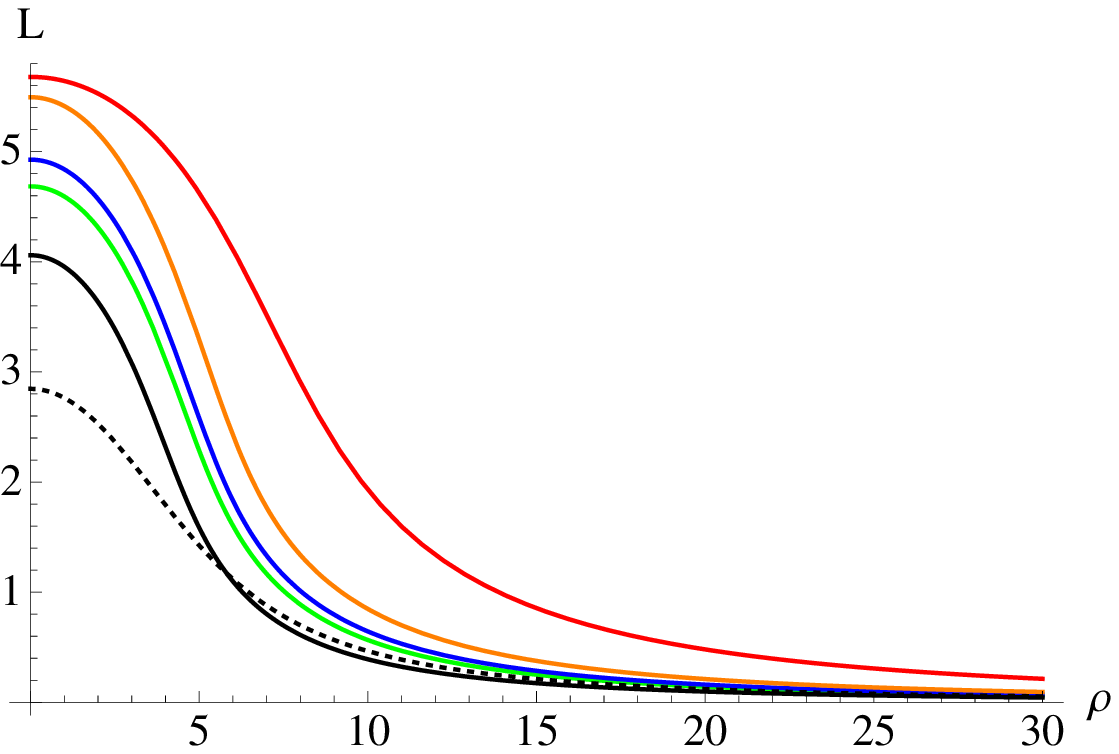}}
      \begin{tabular}{|c|c|c|c|}
      \hline  & $\G$ & $A$ & $\l$ \\
      \hline Black & 1 & 3 & 3.240 \\
      \hline Green & 1 & 2 & 4.045 \\
      \hline Blue & 1 & 1.8 & 4.325 \\
      \hline Orange & 1 & 1.5 & 4.940 \\
      \hline Red & 0.5 & 1.8 & 5.882 \\
      \hline Black(Dotted) & B = 35.6 & - & - \\
      \hline
   \end{tabular}
       \caption{
           {\small The coupling flow (\ref{beta}) (top) and 
            the D7 brane embeddings/quark self energy (bottom).
           }
           }\label{Embed}
\end{figure}

We again display the embeddings for some particular cases in Fig
\ref{Embed}. Note that we have chosen parameters here that make
the potential difference between the symmetric and symmetry
broken phases the same in each case. This is
crucial to ensure that we are comparing models that will generate
the same cosmological constant and hence the same rate of
inflation when the quark condensate is zero. The vacuum energy is
given by the DBI action evaluated on the solution. In fact this
energy is formally divergent corresponding to the usual
cosmological constant problem in field theory. As usual we 
subtract the UV component of the energy - we do this by
subtracting the energy of the symmetry breaking, lowest energy
embedding in each case.

The symmetry breaking of these solutions is manifest. The U(1)
symmetry corresponds to rotations of the solution in the $w_5$-$w_6$
plane. An embedding along the axis corresponds to a massless quark
with the symmetry unbroken (this is the configuration that is
preferred at high temperature and it has zero condensate $c$).
The vacuum, curved configurations though map onto that case at large $\rho$
(the UV of the theory) but bend off axis breaking the symmetry in
the IR.

One can interpret the D7 embedding function as the dynamical self
energy of the quark, similar to that emerging from a gap equation 
\cite{Nambu:1961tp}.
The separation of the D7 from the $\rho$ axis is the mass at some
particular energy scale given by $\rho$ - in the ${\cal N}=2$
theory where the embedding is flat the mass is not renormalized,
whilst with the magnetic field or running coupling an IR mass
forms.

\subsection{Approximate Potentials}

\begin{figure}[]
\centering
   {\includegraphics[width=8cm]{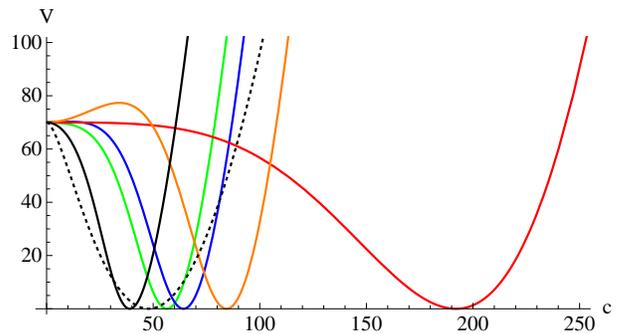}}
\caption{
           {\small The approximate potential, $V \equiv S_{D7}/\wt{T}_7 $,
for the quark condensate $c$,
from the holographic model for the running couplings and D7 embeddings
of Fig \ref{Embed}.}
           }\label{Potential}
\end{figure}

It is natural to want to plot the potential for the quark
condensate using the holographic description. However, this is
somewhat ambiguous. The embedding equation determines the D7
embeddings that correspond to the turning points of such a
potential. In between these points one would need to find the
minimum action configuration that falls off in the UV as
$c/\rho^2$ for arbitrary $c$ and has $L'=0$ in the IR. This
can't be done by a simple numerical shooting because such
configurations are not solutions of the Euler Lagrange equations.
A reasonable way to get qualitative results though is as follows -
we simply take the vacuum embedding solution and plot the
potential as a function of that embedding multiplied by an
arbitrary constant. This gives embeddings for all $L(0)$ and all
values of $c$. The case when we multiply by zero is of course the
central maximum of the wine bottle potential. We will assume that
this form is appropriate for all values of $L(0)$ or $c$. This
then lets us plot the potential for each of our models above - the
potential automatically takes the correct value at the maximum and
minimum. See Fig \ref{Potential}.

Although this procedure is somewhat adhoc it allows a first rough
understanding of the potential shape. We can see that the
potential for the case of a magnetic field induced symmetry
breaking is rather steep around the origin. The step function
ansatz for $\beta$ is gentler there and by decreasing $A$ the
curvature is further reduced. Note that as one decreases $A$ one
must increase $\lambda$ in order to keep the value of the
potential when $c=0$ equal. In the extreme low $A$ limit the
model is characterized by that potential value being much less
than the characteristic scale of the running $\Lambda$. The
quark condensate also grows in this limit.
Comparing the blue curve and the red curve shows that varying
$\Gamma$ also leads to a potential with a flatter and longer
period near the origin.
This is encouraging since for an inflation model we would want the potential to be
as flat and as extended as possible around the origin.

The lowest $A$ curve (Orange) shows a new minimum at zero and an
intermediate maximum in the potential but this is an artefact of
the crude approximation being made here. Were the maximum to
really form then there would be a new solution to the embedding
Euler-Lagrange equation for the D7 brane - we have checked and no
such solution exists. We are simply describing the non-vacuum
configurations incorrectly. Fig \ref{Potential} is we stress just
to show qualitatively that a Higgs-like potential exists in the
theory. Indeed we will see below that the time evolution
of a rolling configuration depends crucially on the behaviour
in the holographic directions of the description of the gauge theory
which these plots are blind to. To do a better job of understanding the dynamics of these
models near the origin we turn to studying the full time dependent
problem of a moving D7 brane.

\section{Holography Of Inflation}

\subsection{Time Dependent Holography}

We will now move on to look at time dependent evolution in the
gauge theory described above. Strictly the AdS/CFT Correspondence
describes our gauge theory in rigid flat space. To include
inflation we will instead assume a background holographic geometry
given by
\begin{eqnarray}
  && ds^2 = {1 \over \guv} \left[{r^2
\over R^2} ( g_{tt}dt^2 + g_{ij} dx^idx^j) \right. \nn \\
 && \qquad \left. + {R^2 \over r^2} \left( d\rho^2 + \rho^2 d\Omega_3^2 + dw_5^2 +
dw_6^2 \right)\right] \ , \label{ametric}
\end{eqnarray}
where $g_{tt} = -1$ and $g_{ij} = a(t)^2 \d_{ij}$.
For this to be valid even as an approximation we must formally
assume that $a(t)$ is growing much slower than the typical time
for equilibration in the ${\cal N}=4$ Yang-Mills plasma. How then
can we hope to study inflation with exponential space-time growth?
The formal answer is that we will study inflation driven by the
flavour sector of our gauge theory. The contribution of the D7
branes to the full vacuum energy of the field theory is $1/N_c$
suppressed relative to the glue dynamics contributions. We will be
able to study the theory where this perturbation to the vacuum
energy drives inflation but formally in a limit where the
inflation is still slow relative to the ${\cal N}=4$ equilibration
time. Our inflation is that relevant for the quark fields. This
provides a formal justification for what follows.

In fact for a true model of inflation one would want exponential
expansion relative to all time scales in the problem. The reader
can therefore choose to view (\ref{ametric}) as a phenomenological
model. In practice for the results we wish to extract - what
dilaton profile will generate slow rolling behaviour - one could
just work in static $x_4$ flat space. However, it is useful to have the
damping term in the scalar equation of motion in (\ref{scalar}) to
remove late time oscillations - this provides another
justification for using (\ref{ametric}).

The DBI action in such a geometry is given by \beq \label{DBI}
 S_{M} = S_{DBI} = \int d^4x \sqrt{-g} \, \call
\eeq
where
\beq
\mathcal{L} = - \wt{T}_7 R^4 \int d \rho \, \b \, \rho^3\sqrt{ {g^{00} \dot{L}^2 \over (\rho^2 + L^2)^2}
 + 1+L'^2 } \label{Lag4D} \eeq
Here $g_{\mu \nu}$
are boundary field theory metric components and $L = \w_5(t,\rho)$ with $\w_6 = 0$.
$L' = \del_\rho L$ and $\dot{L} = \del_t L$.
All variables ($\rho, L, t, \G, \l $) in the integral (\ref{Lag4D}) are rescaled by $R$ and
dimensionless. We will reinstate $R$ when needed.
Note that
the 4D effective action derived from the DBI action can be written
in a covariant form with the correct measure.

The integral form of the 4d Lagrangian (\ref{Lag4D}) may be understood as
the effective Lagrangian of some field theory quantity after integration
over the extra direction $\rho$.
Since the bulk embedding dynamics is closely related to the spontaneous
symmetry breaking in the boundary field theory, it is natural to
associate an effective degree of freedom to an order parameter,
the condensate $c(t)$.
Thus we may consider (\ref{Lag4D}) as an inflaton model of
a composite scalar field $c(t)$ with some potential $V(c)$.
%
%
The quark condensate $c$ can be extracted numerically
from the asymptotic large $\rho$ form of the solution (\ref{asy}).
Even though it's not straightforward to read the form of $V(c)$
from (\ref{Lag4D}) it is not an obstacle to the study of inflation,
since $\ve$ and $p$  can be
computed from (\ref{Lag4D}) and we can 
apply the inflation condition (\ref{negativep}). 
Furthermore our potential for the condensate has its origin in (\ref{Lag4D}),
so is determined in principle and not put in by hand.

We can obtain  $\ve$ and $p$
by computing the expectation value of the stress energy tensor
of the 3+1d field theory from \beq \langle T^{\mu \nu} \rangle = {
2 \over \sqrt{-g}} { \delta S_{M} \over \delta g_{\mu \nu} } \ .
\eeq According to the AdS/CFT Correspondence the gravitational
action is the Masterfield of the field theory
\cite{Malda,Witten:1998qj,Witten:1998b,Gubser:1998bc} with the
boundary values of gravitational fields playing the roles of the
sources. Here we have explicitly shown the behaviour of the
gravity action on the boundary metric so we may simply compute. We
find
\begin{eqnarray}
  && \varepsilon(t) =  \wt{T}_7 R^4 \int d \rho { \rho^3 \b \frac{ 1
+ L^{'2}}{\sqrt{1 +L'^2- {\dot{L}^2\over (\rho^2 + L^2)^2}}}} - \varepsilon_0 \qquad \label{veD7} \\
  && p(t) = - \wt{T}_7 R^4 \int d \rho  \rho^3 \b \sqrt{1
+ L^{'2}- {\dot{L}^2\over (\rho^2 + L^2)^2} }  - p_0  \qquad \label{pD7}
\end{eqnarray}
where we have renormalized by subtracting $\varepsilon_0,p_0$, which
are the values of the integrals with the asymptotic static symmetry
breaking solution for $L$, say $L_s$.
Examples of $L_s$ are shown in Fig \ref{Embed}.
There is no explicit $a(t)$ dependence in $\ve$ and $p$, but its information is encoded in $L(t,\rho)$
since $L(t,\rho)$ is a solution of equation of motion of
(\ref{DBI}) which includes $a(t)$ in $\sqrt{-g}$.

Note that $p(0) = -\ve(0) $ if $\dot{L} = 0$.
So if an initial condition has $\dot{L} \sim 0$ then the Universe will start
inflating ($p < -\ve/3$) regardless of an initial $L'$.
However we know that eventually the embedding will asymptote to $L_s$ because
of the wine bottle shaped potential, which implies $\ve \ra 0$ and $p \ra 0$.
Thus we expect $\ve$ is decreasing and $p$ is increasing with time.
If $p$ increases faster than $\ve$ then there will be a time when
$\ve + 3p = 0 $, ending the inflation.
How long it takes depends on
the initial configuration $L(0,\rho)$ and the parameters of $\b$, the
dilaton profile.
According to small field inflation models  
it is natural to start with $L=0$, which is a local maximum configuration
and the symmetric minimum at high temperature.
Then the inflation time or the number of e-folds ($N_e$)
can be studied as a function of coupling $\b$.
In our setup inflation always happens to some degree and the issue is
the amount of inflation.

Since we know $\ve$ and $p$ we can proceed with the Friedmann (\ref{Friedmann})
and fluid (\ref{Fluid}) equations.
Let's start with the Friedman equation.
\begin{eqnarray}
  H^2 = \frac{1}{3}\ve \ ,
\end{eqnarray}
where $\ve$ is (\ref{veD7}).
This is very complicated.
$\ve$ has $H(t)$ in $a(t)$ which is enters in
the equation of motion for $L$.
Also the Friedmann equation is coupled to the fluid
equation.

To make the computation tractable we start with the zeroth order
slow rolling approximation by assuming $\ve$ is almost constant for a
long enough time, namely $\ve = V_0$.
It gives us a simple solution for $a(t)$
\begin{eqnarray}
 a(t) = e^{\sqrt{\frac{V_0}{3}}t} \label{att} \ ,
\end{eqnarray}
where we set $a(0) = 1$.
This solution must be consistent with the fluid equation,
which is equivalent to the equation of motion for $L$.
Thus we will solve the equation of motion resulting from
(\ref{DBI}) with a constant $V_0$ (\ref{att})
and then plug the solution back into (\ref{veD7}).
If the calculated $\ve$ changes slowly enough (slow rolling),
then our solution is self-consistent.
In our numerical computation $\ve$ and $p$ are always rescaled as
$\ve \ra \ve \wt{T}_7 R^4$ and $p \ra p \wt{T}_7 R^4 $ so are dimensionless.

%

\subsection{Rolling in a B field}

As a first example of our formalism we will consider the case of
magnetic field induced symmetry breaking. We do not expect such a
model to be well suited to a realistic inflation model
because the typical curvature of the
potential (Fig \ref{Potential}) for the condensate $c$ is large.
There is no particular fine tuning in the model.
We will be able to track the
time evolution of the brane configuration though from a symmetric
to a symmetry breaking vacuum.

We will first study the time dependence of the model in a
constantly inflating Universe \ie with fixed cosmological constant
$H$. In reality $H$ should be determined by the D7
energy density through the evolution ($H$ depends on the
choices of $N_c$ and the 'tHooft coupling - we will show
some generic numerical results). In the early inflating stages of
the evolution we are most interested in, assuming a constant
$H$ is sound. For the evolution near the true vacuum $H$
should go to zero and the brane would be expected to oscillate
about the true vacuum configuration.

Most of the solutions we will show will remain highly damped
in the late time regime which allows us to
numerically test that the configuration indeed ends on the static
vacuum D7 embedding.  We will though show some
results for  $H=0$, where those oscillations are observable,
shortly. The high damping is due to our unphysical
constant $H$ at all time. In a full computation it would be time dependent
and vanish at late time. If we could solve with a time dependent $H$
we would find the oscillation around the true vacuum at late time
as hinted in the case $H=0$.

In particular we will start with the initial conditions
\beq L(\rho, t=0) = 0, \hspace{1cm} \dot{L}(\rho,t=0) = v e^{-
\rho^2} \ , \label{speed} \eeq
where $\dot{L} \equiv \del_t L$. The initial speed and $\rho$
dependence of this ansatz is not picked for any deep reason but is
just illustrative of some initial condition that initiates the
roll down to the potential minimum. We have checked that none of
our results are qualitatively changed by varying for example the
width of this initial condition. We will typically pick $v$ to be
very small so that the roll time from the peak of the potential is
quite long. This also ensures that inflation happens at early times.
The early time inflationary period in the plots below are the
result of this fine tuned small initial condition and the large damping term
and not a sign that
the potential is particularly flat near the origin.

A slow early roll will mean we can study the dynamics more easily
numerically and understand better how to lengthen that early roll
period. Through the roll we must ensure that $L'(0,t)=0$ and that
$L(\infty, t) = 0$. The evolution can in fact be followed with
these boundary conditions using the in built numerical partial
differential equation solver in Mathematica.
\begin{figure}[]
\centering
   {\includegraphics[width=7cm]{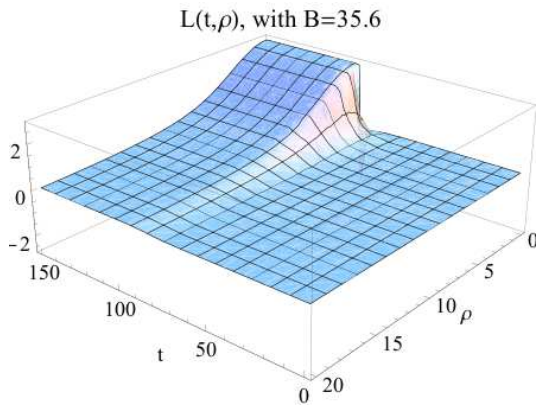}}
\caption{
           {\small We show the time evolution of the D7 brane with
$H=\sqrt{70/3}, B=35.6$ and initial velocity parameter $v=0.00001$.
The transition from the flat embedding (chirally symmetric phase) to the
curved embedding (chirally broken phase)is apparent. }
           }\label{Broll}
\end{figure}

In Fig \ref{Broll} we show a sample plot of the numerical
evolution. The figure is for an inflating Universe with
$H=\sqrt{70/3}, B=35.6$. We show a three dimensional plot for
the time evolution of the D7 embedding - at early times the D7 is
flat, $L=0,$ but as soon as the kinetic energy of the brane begins
to grow as it experiences the curvature of the potential, it
rapidly transitions to the vacuum embedding with symmetry
breaking.
\begin{figure}[]
\centering
\includegraphics[width=7cm]{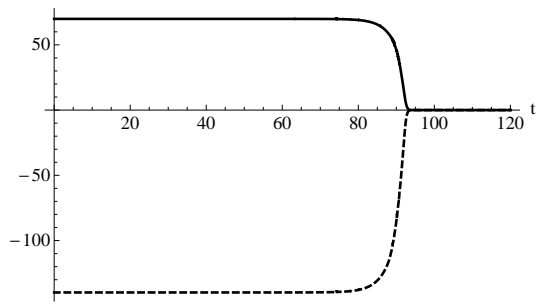}
\caption{
           {\small $\ve+3p$(dotted) and $\ve$(solid) plotted against time for the
time evolution of the configuration shown in Fig \ref{Broll}. }
           }\label{Benergy}
\end{figure}

Fig \ref{Benergy} is a plot of $\ve + 3p$  and $\ve$ versus time.
$\ve(t) \approx -p(t)$ at all time.
$\ve$ shows a plateau until $t\sim 80$ and changes abruptly, while
$\ve+3p$ is negative.
Thus our slow roll approximation is valid up to $t \sim 80$.
Even though the solution after $t\sim 80$ is beyond our approximation
we read off the inflation ending time as the time when $\ve+3p=0$
as an estimate, which is $t_e \sim 93$.
The end time will depends on $B$ or $H$, which are related.

\begin{figure}[]
\centering
   {\includegraphics[width=7cm]{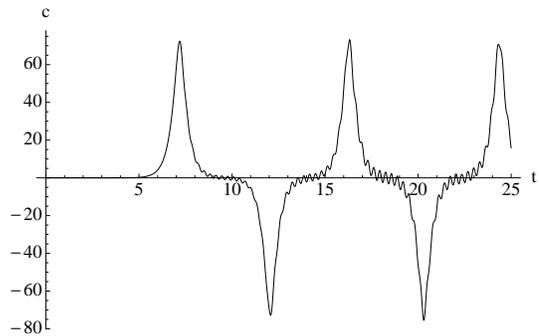}}
\caption{
           {\small We show the time evolution of the quark condensate with
$H=0, B=30$ and initial velocity parameter $v=0.00001$. The
D7 brane oscillates from the flat embedding (chirally symmetric
phase)
 to the
curved embedding (chirally broken phase) above and below the axis.  }
         }  \label{kapzero}
\end{figure}

We stress again that we have taken the Hubble parameter
$H$ constant through the brane motion in the results just
presented. With this unphysical late time damping it can be seen
that the solutions precisely match on to the time independent
vacuum configuration at large time. To show the oscillatory
behaviour one expects when the Hubble parameter is not
present we can also solve for a similar configuration with $H=0$.
We show a plot in Fig \ref{kapzero} of the
time dependence of the quark condensate $c$ in such a scenario -
$c$ is extracted from the large $\rho$ dependence of the solution
through (\ref{asy}). With no damping at all the D7 moves to
approximately the vacuum configuration, then overshoots, returns
to the flat embedding before moving below the axis, etc. The
oscillatory behaviour is clear and can be followed through many
cycles. 
Note that there are ``fine wrinkles'' near $c=0$ in this plot. 
These are due to a peculiarity of the magnetic field induced symmetry breaking -
in particular there are an infinite set of meta-stable vacua near $c=0$ 
in this theory as explored in ~\cite{Spirals}. We see their influence on this motion
although they play no particular role in our analysis here. Our
phenomenological dilaton profiles below do not generate such structure.

Let us briefly return to compare these simulations to what would
be needed for inflation in our Universe. The energy density (\ref{DBI}) is naively
\begin{eqnarray}
  \ve = \tilde{T}_7 R^4\mathbb{E}(B)  , 
\end{eqnarray}
where  $\mathbb{E}(B)$ is a number obtained by the numerical integration in
(\ref{Lag4D}). Note the choice of the magnetic field here introduced the 
intrinsic scale of symmetry breaking.

$\ve$ is also a measure of the symmetry breaking scale of the theory and in flat
space we could set it to one as the defining scale in the problem. When we
include the inflationary dynamics through the damping term in (\ref{Lag4D})
the Planck scale also enters. The ratio of the Planck scale to the fourth root of 
$\ve$ is a free parameter. 
We will associate the Planck mass with 
some distance in the holographic $\rho$ direction, $\gamma_{p}$.
The physical Planck mass in
the coordinates with which we compute (rescaled by $R$ as discussed under (\ref{Lag4D}))
is thus
\beq m_{P} =  {\gamma_{p} \over R} \ . \eeq
It is then sensible to express the energy density in Planck units so
\beq \ve = \ve_0 m_{P}^4 \ ,\eeq
where
\begin{eqnarray}
  \ve_0 ~ \equiv ~\mathbb{E}(B) \frac{1}{2(2\pi)^5 
g_{\mathrm{UV}}^2 \g_p^4} 
\frac{R^8}{\a'^4}~  = ~ \mathbb{E}(B) \frac{2\l_{tH} N}{(2\pi)^3 
 \g_p^4}\ ,  \label{aargh}
\end{eqnarray}
with $\l_{tH} \equiv g_{UV}^2 N$. Note this is $\alpha'$ independent as it should be since $R^8/\a'^4= 
(4 \pi g_{UV}^2 N)^2$.

The roll times $\tilde{t}_e$ we have computed are in dimensionless units and should be 
written in Planckian units too. 
\beq t_e m_P = {R \tilde{t}_e m_p} = \gamma_p \tilde{t}_e  \ . \eeq
The scale factor in the metric is therefore given by
\begin{eqnarray} \label{at}
  a(t) = e^{Ht} = e^{\sqrt{\frac{8\pi \ve_0}{3}} \g_p \tilde{t}_e} \ .
\end{eqnarray}
Our simulations have been with $\sqrt{\frac{8\pi \ve_0}{3}} \g_p  =\sqrt{70/3}$.
Note one can realize this for any value of $\ve_0$ - once $\ve_0$ is fixed
we can choose $\gamma_p$ by setting the gauge theory parameters such as 
$\l_{tH}$ and  $N_c$.

We conclude this section with the comment that it is quite remarkable
that we can compute so straight forwardly the
time evolution of a strongly coupled gauge theory!

\subsection{Towards an Inflationary Dilaton Profile} \label{ETR}

Our next goal is to look to extend the period of
slow roll in the holographic models by phenomenologically
changing the dilaton profile or gauge coupling's running. This is the equivalent
of making the usual inflaton potential flatter around the origin.

\begin{figure}[]
\centering
   {\includegraphics[width=7cm]{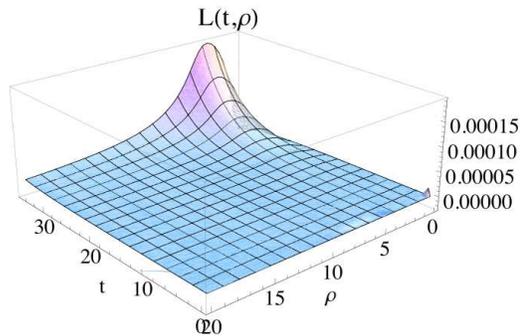}}
\caption{
           {\small The early time evolution of the D7 brane in the
presence of a dilaton step with $A=3$, $\lambda=3.24$, $\Gamma=1$
and for the intial condition $v=0.00001$. $H=\sqrt{70/3}$. The acceleration occurs
around the step. }
           }\label{earlytime}
\end{figure}

\begin{figure}[]
\centering
   {\includegraphics[width=7cm]{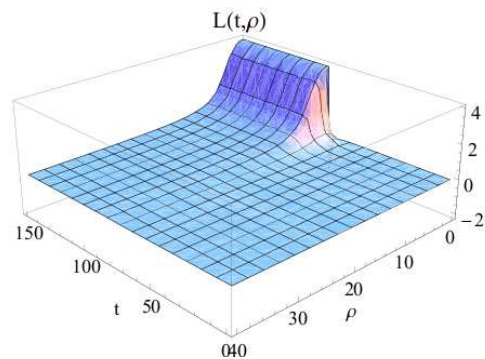}}
\caption{
           {\small The full time evolution of the D7 brane in the
presence of a dilaton step with $A=3$, $\lambda=3.24$, $\Gamma=1$
and for the initial condition $v=0.00001$. $H=\sqrt{70/3}$.  }}
\label{step}
\end{figure}

To
examine this analytically let us first consider the roll
in static space $H=0$. We linearize
the equation of motion about the initial symmetry preserving configuration
$L=0$
\begin{eqnarray}
  \ddot{L} = 3\rho^3L' + \rho^4 L'' + \rho^4 \frac{\b_0'}{\b_0} L' -
  \rho^3 \frac{\b_0'}{\b_0} L   , \label{deracc}
\end{eqnarray}
where $\b_0$ is the coupling at $L=0$ (for the step function form
of the running we have $\, \b_0= A+1-A\tanh[ \G(\rho-\l)]$ and
$\b_0' = - A\,\G\, \mathrm{sech}[\G(\l-\rho)]^2$). If we consider
a truly flat configuration so $L'= L^{''}=0$ then only the last
term on the right hand side contributes. Clearly the acceleration
of the brane is localized around the point in $\rho$ where there
is a significant change in the coupling value. We show that this
is indeed how the motion proceeds in Fig \ref{earlytime} where we
show a plot of the early time evolution of the brane in the
presence of a step in the value of the dilaton. A bump grows
around the point in $\rho$ where the dilaton changes. We also plot
the full evolution with a cosmological damping parameter
$H=\sqrt{70/3}$ showing the configuration move to the true
vacuum in Fig \ref{step}.

For the step configuration with large $\Gamma$, so we can
consider it to be a sharp step, we have
\begin{eqnarray}
  \ddot{L}(\rho) \sim \l^3 \frac{A\,\G}{A+1}L(\l)\d(\rho-\l) \ ,
\end{eqnarray}
naively lowering $A$ or $\Gamma$ will reduce the acceleration. In
fact though to keep the difference in the potential between the
symmetric and symmetry breaking configurations equal as we reduce
$A$ or $\Gamma$ we must increase $\lambda$. For the
numerical values in Fig \ref{Potential}, increasing $\l$ wins so
$\frac{A \Gamma}{1+A} \l^3$ increases when $A$ decreases.

This analysis is overly naive though because as the D7 brane
evolves, as shown in Fig \ref{earlytime}, $L'$ and $L^{''}$ will
become important. Around the localized peak of the bump where
$L'=0$ $L^{''}$ is negative and the equation of motion will be
\begin{eqnarray} \label{ldp}
  \ddot{L}(\rho) \sim \left(\l^3 \frac{A\,\G}{A+1}L(\l)+\l^4 L''(\l)\right)\d(\rho-\l) \ .
\end{eqnarray}
Since $L^{''}$ is negative the latter term will slow the
acceleration in the $A \rightarrow 0, \lambda \rightarrow \infty$
limit. Thus we conclude that for models with fixed $H$ those
with a larger value of $\lambda$ will provide a longer period of inflation.
We want to stress that this conclusion is completely dependent on the
holographic description through evolution associated with the $\rho$
direction - we are learning about the role of the strong interaction dynamics.
One certainly
can not, for example, deduce the motion from the simplistic
approximate potentials we displayed in Fig \ref{Potential}.
More generally we note that the holographic
dependence on $\rho$ introduces considerably more complication to
the evolution and explicit simulation is required.

It is clear that if we wish to prolong the early time roll period we
need to reduce the rate of change of the dilaton and push $\lambda$ far above
the vacuum energy $H$. There are two ways we
can do that within our ansatz (\ref{coupling}) - we can reduce $A$ for a fixed $\G$
so that the step is smaller. We would expect this to
increase the roll time.
Also for fixed $A$ we can decrease $\Gamma$ so the change in
$A$ occurs over a larger $\rho$ range. This reduces the dilaton derivative but spreads the
region of $\rho$ over which the change is occurring so there could be no net
change to the total rate of acceleration - we will need to test this case numerically.

\begin{figure}[]
\centering
   {\includegraphics[width=7cm]{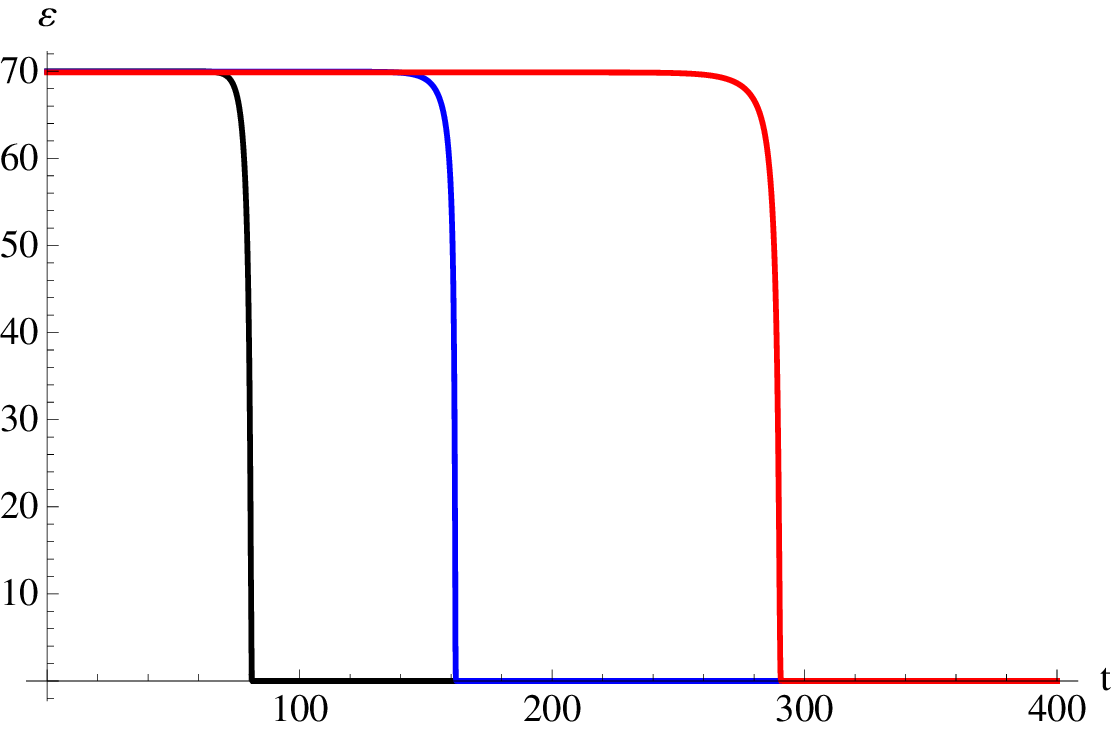}}
\caption{
           {\small Plots of  $\ve$ vs $t$ for the three configurations 
                   [$A=3$, $\Gamma=1$, $\l =3.24$], [$A=1.8$, $\Gamma=1$, $\l =4.325$] and
                    [$A=1.8$, $\Gamma=0.5$, $\l =5.882$]. $v=0.00001$. $H=\sqrt{70/3}$.}
           }\label{accel}
\end{figure}

In Fig \ref{accel} we plot $\ve$ for the three different sets of parameters in $\beta$.
Like Fig $\ref{Benergy}$, $\ve+3p$ is always negative before it suddenly vanishes.
Even though our computation is not valid when $\ve$ starts changing
fast we choose to take the time when $\ve+3p$ vanish as the end time of
inflation, $t_e$. Since our purpose is to compare the inflation time
for different dilaton profiles qualitatively, this approximation will not make a difference.

\begin{figure}[]
\centering
   {\includegraphics[width=7cm]{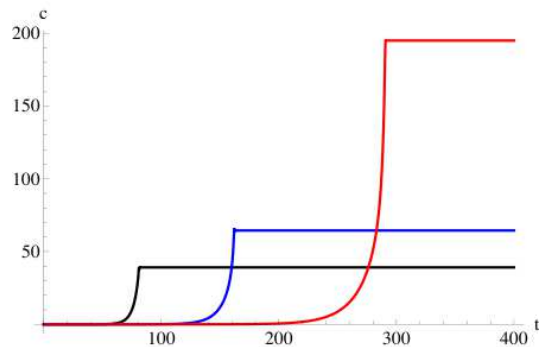}}
\caption{
           {\small The condensate $c$ vs $t$,
             for the three configurations
                   [$A=3$, $\Gamma=1$, $\l =3.24$], [$A=1.8$, $\Gamma=1$, $\l =4.325$] and
                    [$A=1.8$, $\Gamma=0.5$, $\l =5.882$]. $v=0.00001$. $\kappa=\sqrt{70/3}$. }
           }\label{compare}
\end{figure}

We will now make a comparison of the roll time for a number of different
dilaton step profiles. In each case the difference in vacuum energy between the
symmetric and vacuum symmetry breaking configuration is the same. We also use the
same initial velocity perturbation for the D7 brane (\ref{speed}).
As an example of the differences
we plot the energy density $\ve$ against time in Fig \ref{accel} and the condensate $c$  against
time for three configurations in Fig \ref{compare}.
First compare the curves for the dilaton parameters [$A=3$, $\Gamma=1$, $\l =3.24$] and
[$A=1.8$, $\Gamma=1$, $\l =4.325$]. Decreasing $A$ indeed increases the time the configuration
takes to reach the tipping point to the true vacuum. Next we can change $\Gamma$ to try to
further increase the roll time - the final curve is for the configuration [$A=1.8$, $\Gamma=0.5$, $\l =5.882$]
and indeed we find a further lengthening of the inflationary period.

\begin{figure}[]
\centering
   {\includegraphics[width=7cm]{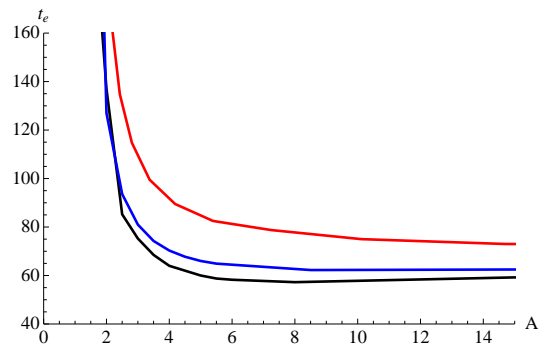}}
\caption{
           {\small $t_e$ against $A$ for a
sequence of values of $\Gamma$s: $1.5$(Black), $1$(Blue), $0.5$(Red). $v=0.00001$. $H=\sqrt{70/3}$.}
           }\label{compare2}
\end{figure}

Finally we show these trends in more detail in Fig \ref{compare2} where we plot $t_e$ against $A$ for a
sequence of values of $\Gamma$ ($1.5$, $1$, $0.5$). There is a clear trend for decreasing both $A$
and $\Gamma$ increasing
the roll time. As this roll time increases the parameters become more fine tuned
reflecting the fine tuning we are making in the effective potential for $c$, ie the usual
fine tuning in inflation.

\begin{figure}[]
\centering
   {\includegraphics[width=7cm]{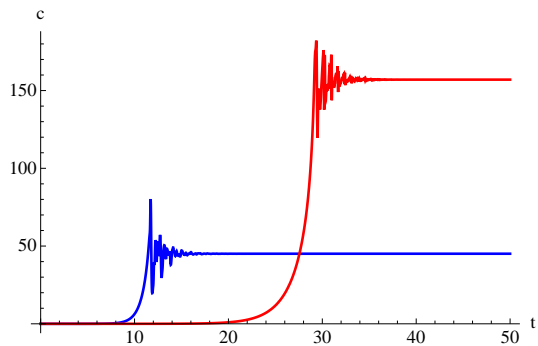}}
\caption{
           {\small $t_e$ against $A$ for two cases
sequence of values. Blue: [$A=27.3$, $\Gamma=0.5$, $\l =0.25$],
Red: [$A=2$, $\Gamma=0.5$, $\l =5.326$]. $v=0.00001$. $H=\sqrt{1/3}$.}
           }\label{compare3}
\end{figure}

This analysis has been performed again in a heavily damped scenario which as discussed 
in (\ref{at}) 
assumes a growing value of $N$ in the gauge theory as the energy density is
reduced relative to the Planck scale. We choose this regime primarily
for computational convenience.
Having high damping removes late time oscillations of the D7 motion. It also
extends the roll time making changes in that roll time more easily apparent.
To demonstrate that the effects we have observed are still present at
lower values of the damping we finally display the behaviour of the condensate
against time for the case $H=\sqrt{1/3}$ and for steps with two different values
of step height $A$ in Fig \ref{compare3}. The late time oscillations are 
now apparent but the increase of the roll time with decreasing $A$ is 
also maintained.

\section{Discussion}

Inspired by small field inflation models, we have used holography to
study out of equilibrium dynamics in a strongly coupled gauge theory with
chiral symmetry breaking. We have shown that we can extract an approximate
effective potential for the quark condensate. We have then been able to
explicitly follow the roll of a configuration from the symmetric potential maximum
to the symmetry breaking potential minimum. We first performed this computation in
a model with chiral symmetry breaking induced by a magnetic field. While
the configuration lies near the potential maximum it generates inflation. Formally
we worked in a probe limit that requires that the rate of inflation is small
relative to the typical relaxation time of the background ${\cal N}=4$ gauge
dynamics.

We then studied a holographic model of a
large $N_c$ gauge theory with an arbitrary running of the gauge coupling. 
It is a model in the spirit of AdS/QCD since we do not backreact 
the geometry to the dilaton profile. 
The dilaton profile we chose to study means the gauge theory 
has a strongly coupled conformal UV
and deforms to an IR conformal theory with a higher value of
the gauge coupling. That change in the coupling triggers chiral symmetry
breaking for the quarks.
We have found that when the change in the coupling is fine tuned to be small
or slow running
the theory can give rise to inflationary dynamics. An important
contribution to this slowing of the roll was made by the D7 dynamics in the
radial direction of the geometry through (\ref{ldp}) - this suggests that 
the strong dynamics of these gauge theories encourages inflation.
Are there broad lessons
to be learnt here about strongly coupled gauge theories or is this an oddity
of a particular AdS set up?

There is a growing belief that many asymptotically free gauge
theories indeed give rise to strongly coupled conformal regimes.
Seiberg's dualities for ${\cal N}=1$ supersymmetric QCD \cite{Seiberg:1994pq}
 were the
first hint - they show that SQCD flows to a non-trivial IR
conformal theory in the range $N_c+1< N_f < 3 N_c$. For much of
that range the UV degrees of freedom are strongly coupled. Near
$N_f=3 N_c$ and at large $N_c$ these phases match onto the
perturbative Banks Zaks fixed points \cite{Banks:1981nn}. Banks Zaks fixed points also
exist in non-supersymmetric theories and it is reasonable to
expect that IR conformal fixed points also exist for a
considerable range of $N_f$ in those theories (see \cite{Appelquist:1996dq}, \cite{Ryttov:2007sr} for some speculation about these theories). Theories
with higher dimension representation matter fields would also be
expected to generate strongly coupled IR fixed points. Recently
there has begun to be lattice simulations of QCD with varying
unquenched quark flavours \cite{Appelquist:2009ty,Appelquist:2007hu} and higher dimensional representation
matter present \cite{DelDebbio:2009fd,DelDebbio:2008zf}
- there is certainly encouragement in these results
for the view that IR conformal theories exist in some of these
cases.

If we believe that such fixed points are fairly common then we can imagine
several ways for how
to construct a theory with the profile for the running coupling we have
studied in this paper. For example, one could begin with an SU($N_c$) theory
with sufficient non-fundamental matter to place it at a strongly coupled
fixed point. If $N_c$ is appropriately large then we can also add fundamental
quark multiplets as a perturbation - their contributions to the beta
function coefficients will be $N_f/N_c$ suppressed and hence they will most
likely generate just a small change in the fixed point's coupling value. The
fundamental quarks though can be used to dial a profile for the running
coupling if they can be sequentially decoupled. For example, if they are
vector-like one could just put in masses to adjust the running at the order
$N_f/N_c$ shifting the IR theory from one conformal point to another. This
realizes the running we described above. It is not obvious how chiral symmetry
breaking might happen in this scenario though. Usually one imagines
an NJL-model-like critical coupling for chiral symmetry breaking \cite{Nambu:1961tp}. One would
need to tune the shift in the coupling to cross that critical value for the
fundamental quarks. Naively higher dimension representation quarks might also
be expected to condense at the same point or at a lower value of the coupling
though.

Another possibility, which is sometimes discussed \cite{Brodsky:2010vh},
is that pure glue gauge theories in fact have an IR
fixed point for the coupling. One could imagine perturbing that fixed point
with some fundamental matter and again, by appropriate decoupling, change the
running to match that we seek. The pure glue fixed point would again need
to lie just above the critical coupling for chiral symmetry breaking.

Finally the other possibility we are led to is walking dynamics   \cite{Holdom:1981rm,Appelquist:1986an}. A theory
with fundamental quarks that is approaching an IR fixed point where the
coupling is tuned
just above the critical coupling might spend many RG decades at strong coupling
but without triggering chiral symmetry breaking before
finally reaching the critical value at a low scale. Equally the coupling might
cross the critical value at a high scale but be so close to the critical
value from above that the quarks' dynamical mass is at a much lower scale
so they don't decouple and the fixed point is maintained.  Such a theory would
naturally realize
both the small $A$ and small $\Gamma$ limit of our coupling ansatz, both of
which led towards inflation. Walking is of course proposed as a solution
of the mini-hierarchy problem in technicolour theories of electroweak
symmetry breaking \cite{Weinberg:1975gm, Susskind:1978ms}. 
It is certainly intriguing if the fine tuning already
used to solve that problem also generates inflationary dynamics.

In conclusion it certainly seems possible that a range of asymptotically
free gauge theories might realize the behaviour we have seen 
and generate effective dynamics that encourages inflation. The
need for intrinsic fine tuning, as usual in small field inflation, seems unavoidable though.

\acknowledgements

The authors thank Beatriz de Carlos, Ingo Kirsch and Jonathan Shock for discussions.
We are grateful for support from STFC. NE and KK thank the ESI, Vienna
for hospitality where some of this work was performed.

\end{document}